\documentclass[aip,jcp,reprint]{revtex4-1}
\usepackage{graphicx}
\usepackage{color}
\usepackage{amsmath}
\usepackage[colorlinks=true, linkcolor=blue, urlcolor=blue, citecolor=blue]{hyperref}

\begin{document}

\title{Sequence dependent aggregation of peptides and fibril formation}

\author{Nguyen Ba Hung}
\affiliation{Institute of Physics, Vietnam Academy of Science and Technology,
10 Dao Tan, Ba Dinh, Ha Noi, Viet Nam}
\affiliation{Graduate University of Science and Technology,
Vietnam Academy of Science and Technology, 18 Hoang Quoc Viet, Cau Giay, Ha Noi, Viet Nam}
\affiliation{Vietnam Military Medical University, 160 Phung Hung, Ha Dong, Ha
Noi, Viet Nam}

\author{Duy-Manh Le}
\affiliation{Institute of Research and Development, Duy Tan University, K7/25
Quang Trung, Da Nang, Viet Nam}

\author{Trinh X. Hoang}
\email{hoang@iop.vast.vn}
\affiliation{Institute of Physics, Vietnam Academy of Science and Technology,
10 Dao Tan, Ba Dinh, Ha Noi, Viet Nam}
\affiliation{Graduate University of Science and Technology,
Vietnam Academy of Science and Technology, 18 Hoang Quoc Viet, Cau Giay, Ha Noi, Viet Nam}

\date{\today}

\begin{abstract}
Deciphering the links between amino acid sequence and amyloid fibril
formation is key for understanding protein misfolding diseases. Here we use
Monte Carlo simulations to study aggregation of short peptides in a
coarse-grained model with hydrophobic-polar (HP) amino acid sequences and
correlated side chain orientations for hydrophobic contacts. A significant
heterogeneity is observed in the aggregate structures and in the thermodynamics
of aggregation for systems of different HP sequences and different number of
peptides.  Fibril-like ordered aggregates are found for several sequences that
contain the common HPH pattern while other sequences may form helix bundles or
disordered aggregates. A wide variation of the aggregation transition
temperatures among
sequences, even among those of the same hydrophobic fraction, indicates that
not all sequences undergo aggregation at a presumable physiological
temperature. The transition is found to be the most cooperative 
for sequences forming fibril-like structures. For a fibril-prone sequence,
it is shown that fibril formation follows the nucleation and growth mechanism.
Interestingly, a binary mixture of peptides of an aggregation-prone and a
non-aggregation-prone sequence shows association and conversion of the latter
to the fibrillar structure. Our study highlights the role of sequence in
selecting fibril-like aggregates and also the impact of structural template on
fibril formation by peptides of unrelated sequences.
\end{abstract}

\maketitle

\section{Introduction}

The phenomenon in which soluble proteins or protein fragments self-assemble
into insoluble aggregates is considered as a fundamental issue of protein
folding with serious impact on human health \cite{ChitiAnnRev2006}.
A predominant class of these aggregates, that have a long straight shape and
are rich in $\beta$-sheets, known as amyloid fibrils, is associated to a range
of debilitating human pathologies, such as Alzeihmer's, Parkinson's, type II
diabetes and transmissible spongiform encephalopathies \cite{Riek2016}.
These fibrils, formed by numerous proteins and peptides
including those unrelated to disease \cite{Jimenez1999}, have
strikingly similar structural features regardless of the amino acid sequence.
An widely adopted view is that the tendency of forming amyloid fibrils is 
a common property of all proteins, supposedly due to their common 
polypeptide backbone \cite{Dobson1999}. 
It has been shown that poly-aminoacids can also form amyloid under
appropriate condition \cite{Fandrich2002}.
However, the propensity of a given polypeptide to form amyloid fibrils as well
as the condition under which they form depends very significantly on its amino
acid sequence showing that the problem is much more complex than it could be
initially thought of but also giving hope for curing amyloid diseases
\cite{Ventura2005}. 

X-ray fiber diffraction data indicate that amyloid fibrils are 
commonly characterized by the cross-$\beta$-sheets with strands running
perpendicularly to the fibril's longitudinal axis \cite{Sunde1997}. The
cross-$\beta$-structures at atomic resolution have been obtained for the
fibrils of a few proteins and protein fragments including those of insulin
\cite{Jimenez2002}, $\beta$-amyloid peptide \cite{Tycko2002}, 
yeast prion protein sup35p \cite{Griffin2007}, HET-s prion
\cite{Meier2008}, and $\alpha$-synuclein \cite{Tuttle2016} by using
cryo-electron microscopy, X-ray and solid-state NMR. It is found that they are
highly ordered and composed of $\beta$-strands of the same segments of
repetitive protein molecules. Between the mated $\beta$-sheets is a complete
dry and complementary packing of amino acid side chains with a well-formed
hydrophobic core \cite{Sawaya2007}.  Even though there are evidence of
polymorphism \cite{Petkova2005} in amyloid fibrils, the observed packing of
side chains in the resolved structures has suggested that the amino acid
sequence dictates much the amyloid fold \cite{Meier2015}, in the same manner as
in protein folding.  

The sequence determinant of amyloid formation has been studied with various
experi-mental
\cite{Hecht1999,Hecht2002,Chiti2002,Ventura2004,delaPaz2002,delaPaz2004,Linquist2009}
and theoretical \cite{Colombo2005,Shea07,Li2010,Abeln2014} approaches.  It has
been shown that the overall hydropho-bicity \cite{Chiti2002} and net charge
\cite{delaPaz2002}
of a peptide, to some extent, may impact the aggregation rate. There are
increasing evidence that the capability of a protein to form
amyloids strongly depends on certain short amino acid stretches in the sequence
\cite{Hecht2002,Chiti2002,Ventura2004}.
To support a proteome-wide search for aggregation-prone peptide segments, a
number of predictors have been made available \cite{Tango,PASTA,Waltz}.
However, the problem still substantially needs better understanding.

In this study, we investigate the selectivity of aggregate structures by the
amino acid sequence and the mechanism of fibril formation by using the tube
model of protein developed by Hoang {\it et al.} \cite{HoangPNAS04}. The
latter is a C$_\alpha$-based model exploiting the tube-like symmetry
\cite{MaritanNature} of a polypeptide chain and geometrical constraints imposed
by hydrogen bonds \cite{BanavarPRE04}. Such symmetry and geometry consideration
leads to a presculpted free energy landscape \cite{HoangPNAS04} with
marginally compact protein-like ground states and low energy minima 
\cite{Trovato2005,Hoang2006}. 
Interestingly, the model also shows a strong tendency of multiple chains
to form amyloid-like aggregates \cite{BanavarPRE04,HoangPNAS06}, similar
to that found in higher resolution models \cite{Nguyen2004,Nguyen2005,Shea07}.
Extensive simulations have been carried out by Auer and coworkers
\cite{Auer2007,AuerPRL08,Auer2010,Auer2011} to study the fibril formation of
12-mer homo-peptides using the tube model with a slightly different
constraint on self-avoidance, showing useful insights on the nucleation
mechanism \cite{Auer2007,AuerPRL08} of fibril formation and 
on the equilibrium conditions between the fibrillar aggregates
and the peptide solution \cite{Auer2010,Auer2011}.
In the present study, we focus on the impact of amino acid sequence on the
aggregation properties in the tube model with a renewed consideration of
hydrophobic interaction. In the original tube model, the latter
was based on an isotropic contact potential between centroids represented by
the C$_\alpha$ atoms. We introduce here a new model for hydrophobic contact
between amino acids that takes into account the side chain orientations. We
find that the latter can direct the interaction between $\beta$-sheets and
promote the formation of ordered and elongated fibril-like aggregates.

We restrict ourself to hydrophobic-polar (HP) sequences and short peptides of
length equal to 8 residues. The consideration of HP sequences is a minimalist
approach in terms of sequence specificity, however is well supported in protein
folding \cite{Dill1997,HoangPNAS06}. Furthermore, the rather simplicity of
amyloid fibril structures also indicates a possible simplification of the amino
acid sequence in determining aggregation properties.  It will be shown that
even with a short length and a few sequences, the systems considered already
exhibit a rich behavior in the morphologies of the aggregates and in their
thermodynamic properties.

For an aggregation-prone sequence, we have studied also the kinetics of fibril
formation. We will try to elucidate the nucleation and growth mechanism of this
process at molecular detail and show evidence of a lag phase. Finally, we have
studied a binary mixture of peptides of two different sequences and find that
amyloid formation can be sequence non-specific, that is a fibril-like template
formed by an aggregation-prone sequence may induce aggregation of a
non-aggregation-prone sequence for a fraction of all peptides. This strong
impact of the template decreases somewhat the sequence determination of
aggregation propensity and suggests that amyloid fibrils could be heterogeneous
in their peptide composition.

\section{Models and methods}

Details of the tube model can be found in Ref. \cite{HoangPNAS04}.  Briefly, it
is a C$_\alpha$-based coarse-grained model, in which the C$_\alpha$ atoms
representing amino acid residues are placed along the axis of a self-avoiding
tube of cross-sectional radius $\Delta=2.5\AA$. The finite thickness of the
tube is imposed by requiring the radius of circle drawn through any three
$C_\alpha$ atoms must be larger than $\Delta$ \cite{Gonzalez,MaritanNature}.
The energy of a given conformation is the sum of the bending energy, hydrogen
bonding energy and hydrophobic interaction energy.  A local bending energy
penalty of $e_R = 0.3 \epsilon >0$, with $\epsilon$ an energy unit, is applied
if the chain local radius of curvature at a given bead is less than 3.2 $\AA$.
Hydrogen bonds between amino acids are required to satisfy a set of distance
and angular constraints on the local properties of the chain as found by a
statistical analysis of protein PDB structures \cite{BanavarPRE04}. Local
hydrogen bond, which is formed by residues separated by three peptide bonds
along the chain, is given an energy of $-\epsilon$, whereas non-local hydrogen
bond is given an energy of $-0.7\epsilon$. Additionally, a cooperative energy
of $-0.3\epsilon$ is given for each pair of hydrogen bonds that are formed by
pairs of consecutive amino acids in the sequence. To avoid spurious effects of
the chain termini, hydrogen bonds involving a terminal residue are given a
reduced energy of $-0.5\epsilon$.  

\begin{figure}
\begin{center}
\includegraphics[width=3.4in]{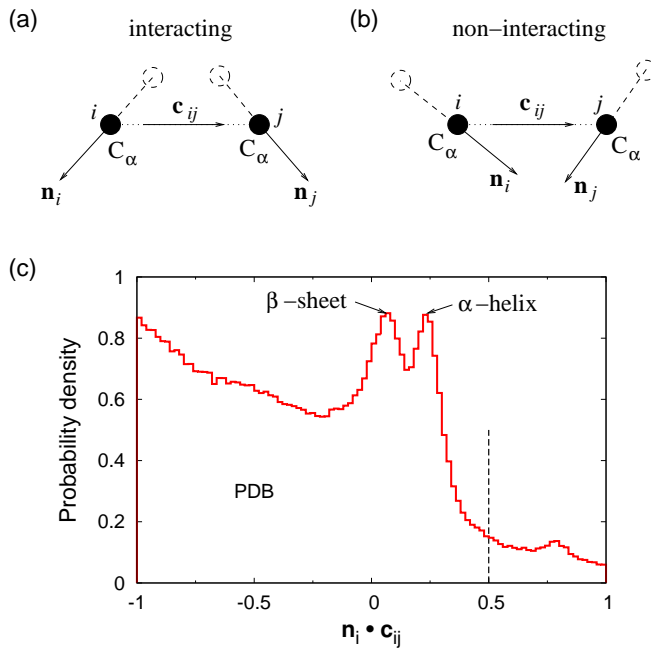}
\end{center}
\caption{\label{fig:sidechain}
(a and b) Model of contact interaction with correlated side chain orientations.
The side chains are assumed to be placed in the inverse direction to the normal
vectors, ${\bf n}_i$ and ${\bf n}_j$, from the $C_\alpha$
atoms. Two amino acids interact if their side chains are oriented towards
each other (a), or do not interact if the side chains are oriented apart
from each other (b).
(c) Histogram of the product ${\bf n}_i \cdot {\bf c}_{ij}$ for
side chain - side chain contacts obtained from 500 filtered PDB structures of
the top500 database. The latter contacts are defined if there are at
least two atoms, separately belonged to the two side chains, found at a
distance less than 1.5 times the sum of their van der Waals radii. The
peaks near the center of the histogram correspond to the contributions of
$\alpha$-helices and $\beta$-sheets as indicated. Vertical dash line indicates
the cut-off used in the model.
}
\end{figure}

Hydrophobic interaction is based on the pairwise contacts between amino acids,
considered to be either hydrophobic (H) or polar (P). It is also assumed that
only contacts between H residues are favorable, and thus the contact energies
of different residues pairing are $e_{HH}=-0.5\epsilon$, and $e_{HP}= e_{PP} =
0$. In the original tube model, a contact is defined if the distance between
two residues is less than 7.5 $\AA$. In the present study, we apply an
additional constraint on hydrophobic contact by taking into account the side
chain orientation \cite{HungJPCS2015} (Fig. \ref{fig:sidechain}a,b). The latter
are approximately given by the inverse direction to the normal
vector \cite{BanavarPNAS09} at the chain's local position. The new constraint
requires that two residues $i$ and $j$ make a hydrophobic contact if ${\bf n}_i
\cdot {\bf
c}_{ij} < 0.5$ and ${\bf n}_j \cdot {\bf c}_{ji} < 0.5$ where ${\bf n}_i$ and
${\bf n}_j$ are the normal vectors of the Frenet frames associated with bead $i$
and $j$, respectively; ${\bf c}_{ij}$ is an unit vector pointing from bead $i$
to bead $j$; and ${\bf c}_{ji}=-{\bf c}_{ij}$. These vectors are given by
\begin{equation}
{\bf n}_i = \frac{{\bf r}_{i-1}+{\bf r}_{i+1} - 2 {\bf r}_i}
{|{\bf r}_{i-1}+{\bf r}_{i+1} - 2 {\bf r}_i|} \ ,
\end{equation}
and
\begin{equation}
{\bf c}_{ij} = \frac{{\bf r}_j - {\bf r}_i}{|{\bf r}_j - {\bf r}_i|} \ ,
\end{equation}
where ${\bf r}_i$ is the position of bead $i$.  The new constraint is in
accordance with the statistics drawn from an analysis of PDB structures (Fig.
\ref{fig:sidechain}b).

We consider 12 HP sequences of length $N=8$ as given in Table I. The
sequences, denoted as S1 through S12, are selected in such a way that they
contain only 2 or 3 H residues, corresponding to hydrophobic fraction of 25\%
and 37.5\%, respectively. We have chosen sequences that are symmetric as
much as possible from the two ends having in mind that the relative positions
of the H residues are more important than their absolute positions in the
sequence. One characterization of these relative positions is the minimum
separation between two consecutive H residues given by the parameter $s$ in
Table I.

\begin{table}
\centering
\caption{\label{hpseq}
HP sequences of amino acids of peptides considered in present study (H --
hydrophobic, P -- polar). The parameter $s$ denotes the minimal sequence
separation between two consecutive H amino acids.}
\begin{ruledtabular}
\begin{tabular}{llr}
{\bf Sequence name} & {\bf Sequence} & {\bf s}\\
S1 & P P P H H P P P & 1\\
S2 & P P H P H P P P & 2\\
S3 & P P H P P H P P & 3\\
S4 & P H P P P H P P & 4\\
S5 & P H P P P P H P & 5\\
S6 & H P P P P P H P & 6\\
S7 & H P P P P P P H & 7\\
S8 & P P H H H P P P & 1\\
S9 & P P H P H H  P P & 1\\
S10 & P H P P H H P P & 1\\
S11 & P H P H P H P P & 2\\
S12 & P H P P H P H P & 2\\
\end{tabular}
\end{ruledtabular}
\end{table}

We will study systems of $M$ peptides
in a cubic box of size $L$ with periodic boundary conditions.
For a given peptide concentration $c$,
the box size $L$ is calculated depending on $M$ as $L=(M/c)^{1/3}$.
For example, for $c=1$ mM (millimolar) and $M=10$ one gets $L=255.15\ \AA$. 
Parallel tempering \cite{Swendsen1986} Monte Carlo schemes with 16-24 replicas
at different temperatures are employed for obtaining the ground state
and equilibrium characteristics. For each replica, the simulation is carried
out with pivot, crankshaft and translation moves and with the
Metropolis algorithm for move acceptance at its own temperature $T_i$. A replica
exchange attempt is made every 10 MC sweeps (one sweep corresponds to a number
of move attempts equal to the number of residues). The exchange of replicas $i$
and $j$ is accepted with a probability $p=\min\{ 1,
\exp[(\beta_i-\beta_j)(E_i - E_j)] \}$, where $\beta=(k_B T)^{-1}$ is
the inverse temperature, $k_B$ is the Boltzmann
constant, and $E_i$ and $E_j$ are the energies of the replicas at the time of
the exchange. 

The temperature range in parallel tempering simulations are chosen such that it
covers the transition from a gas phase of separated peptides at a high
temperature to the condensed phase of the aggregates at a low temperature. The 
replica temperatures are chosen such that acceptance rates of replica exchanges
for neighboring temperatures are significant, of at least about 20\%.
Practically, one needs to change the set of temperatures several times in such a
way that there are more temperatures near the specific heat's peak, where the
energy fluctuation is large. 
For example, for sequence S2 with $M=10$, the final set of temperatures 
for 20 replicas 
is \{0.15, 0.16, 0.17, 0.18, 0.19, 0.20, 0.21, 0.212, 0.214, 0.216,
0.218, 0.22, 0.222, 0.224, 0.226, 0.228, 0.23, 0.24, 0.25, 0.26\} in units of
$\epsilon/k_B$.
The number of Monte Carlo attempted moves is of
the order of $10^9$ per replica. The weighted multiple histogram technique
\cite{Ferrenberg1989} is employed for the calculation of equilibrium properties
such as the specific heat and the effective free energy.

For studying the kinetics of fibril growth,
we carry out multiple independent
Monte Carlo simulations that start from random configurations of dispersed
monomers. These initial configurations are equilibrated at a high temperature
before being used. We are interested in three quantities: the number of
aggregates, the maximum size of the aggregates, and the number of peptides in
$\beta$-sheet conformation during the time evolution. A peptide is said to be
in a $\beta$-sheet conformation if it forms at least 4 consecutive hydrogen
bonds with another peptide.

\section{Results}

\subsection{Sequence dependence of aggregate structures}

\begin{figure}
\includegraphics[width=3.4in]{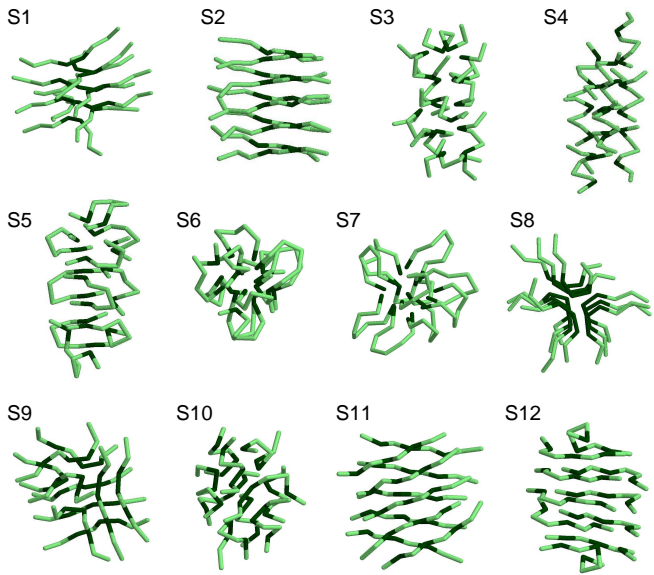}
\caption{\label{fig:galm10} 
Ground state conformations obtained by the simulations for systems of $M=10$
peptides for 10 HP sequences (S1--S10) as given in Table I.
The hydrophobic (polar) residues are shown in dark (light) green color.
}
\end{figure}

We first study the dependence of aggregate structure on the amino acid
sequence for systems of $M=10$ identical peptides at a fixed concentration
of 1 mM. Fig. \ref{fig:galm10} shows that the lowest energy conformation
obtained in the simulations, supposed to be the ground state of a given system,
strongly depends on the sequence. Two sequences, S2 and S11, form a double
layer $\beta$-sheet structure with characteristics similar to that of a
cross-$\beta$ structure. In these structures, an axis of the aggregate
approximately perpendicular to the $\beta$-strands can be drawn. A similar
structure but less fibril-like is also found for sequence S12 with some parts
that are non-$\beta$-sheet. Both sequences S3 and S4 form a $\alpha$-helix
bundle. The helix bundle of sequence S4 however is more ordered and has an
approximate cylinder shape, in which the $\alpha$-helices are almost parallel
to each other. This type of aggregate is akin to non-amyloid filaments formed
by globular proteins such as the actin filament \cite{Hanson1963}.
Other sequences form some sorts of disordered aggregates. In these
disordered structures one may also find a significant amount of
$\beta$-sheets. 
In our model, residues participating in consecutive local and non-local
hydrogen bonds are identified as forming $\alpha$-helix and $\beta$-sheet,
respectively \cite{HoangPNAS04}.

The role of hydrophobic residues in aggregation can be figured out from the
structures of the aggregates. In all cases, one finds the presence of a
well-formed hydrophobic core with the putative hydrophobic side chains oriented
inwards to the body of the aggregate. The packing of hydrophobic side chains is
best observed for sequences S2 and S11, for which the hydrophobic residues are
aligned within each $\beta$-sheet and the hydrophobic side chains from the
two $\beta$-sheets are facing each other.  This packing is possible due to the
HPH pattern in these sequences which position the hydrophobic side chains on
one side of each $\beta$-sheet. An alignment of hydrophobic residues is also
seen for sequence S12 due to the HPH segment of this sequence. In the aggregate
of sequences S4, which is a helix bundle, the hydrophobic side chains are
gathered along the bundle axis, thanks to to the alignment of hydrophobic side
chains along one side of each $\alpha$-helix.  This alignment is due to the
HPPPH pattern in the S4 sequence. On the other hand, the S3 sequence with the
HPPH pattern also forms a helix but the hydrophobic side chains are not well
aligned in the helix, leading to a less ordered aggregate.

\begin{figure}
\center
\includegraphics[width=3.4in]{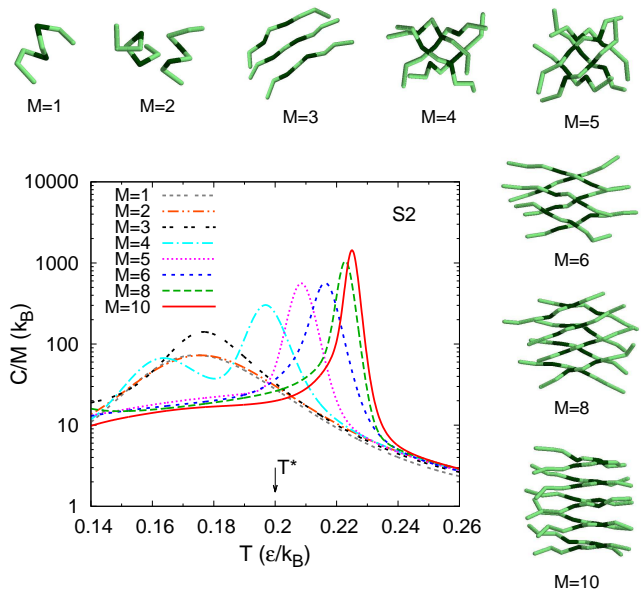}
\caption{\label{fig:s2}
Temperature dependence of the specific heat $C$ per molecule for sequence S2
systems with the number of chains $M$ equal to 1, 2, 3, 4, 5, 6, 8 and 10 as
indicated. The simulation box size $L$ is increased on increasing $M$
such that the peptide concentration is constant and equal to 1 mM.
The conformations shown are ground state conformations obtained by
the simulations for the systems considered. The position of a putative
physiological temperature, $T^*$, is indicated.
}
\end{figure}

\begin{figure}
\center
\includegraphics[width=3.4in]{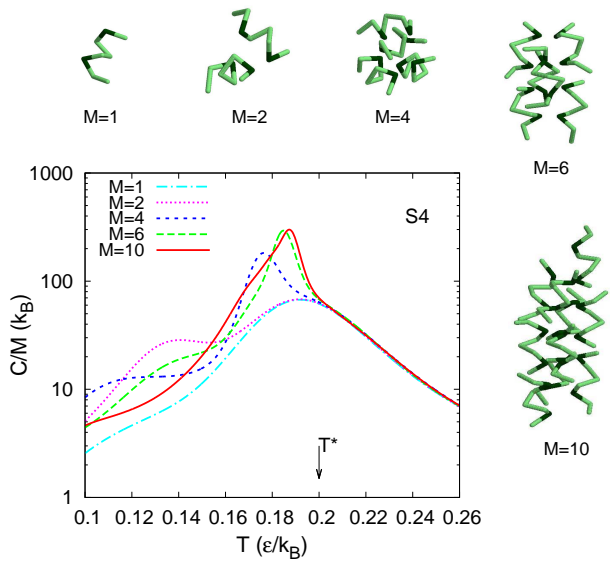}
\caption{\label{fig:s4}
Same as Fig. \ref{fig:s2} but for sequence S4 systems at 1 mM concentration. 
For clarity, the system sizes shown are fewer than for sequences S2.
}
\end{figure}

The structure of the aggregate also depends on the number of chains $M$. In
Fig. \ref{fig:s2} and Fig. \ref{fig:s4}, the ground states for $M$ varying
between 1 and 10 are shown for sequence S2 and S4, respectively.
Interestingly, for sequence S2 (Fig.  \ref{fig:s2}) as $M$ increases one sees
transitions from single helix to two-helix bundle, then to single $\beta$-sheet
($M=3$) and to double $\beta$-sheets ($M\geq 4$). One can also notice that as
$M$ increases the $\beta$-sheet aggregates become more ordered and more
fibril-like as their $\beta$-strands become more parallel. For sequences S4
(Fig. \ref{fig:s4}), only helix bundles are formed for all $M>1$, but the
bundle also becomes more ordered as $M$ increases. Thus, the increasing
orderness with the system size is observed for both $\beta$-sheet and
$\alpha$-helical aggregates.

\subsection{Thermodynamics of aggregation}

It can be expected that the thermodynamics of aggregation depend on the
aggregate structure due to distinct contributions of intermolecular and
intramolecular interactions in different structures. Furthermore, the
formations of ordered and non-ordered aggregates can be different from the
perspective of a phase transition. We will consider the the system's specific
heat, $C$, for the analysis of the thermodynamics. 
We are particularly
interested in the temperature of the main peak of the specific heat,
$T_\mathrm{peak}$, and the peak height, $C_\mathrm{peak}$. 
$T_\mathrm{peak}$ corresponds to the aggregation transition temperature. 
Higher $T_\mathrm{peak}$ means a more stable aggregate, whereas 
higher $C_\mathrm{peak}$ indicates that the aggregation transition
is more cooperative \cite{Kaya2000}.
For all multi-peptide systems considered, it is found that the energy
distribution at $T_\mathrm{peak}$ has a bimodal shape, suggesting that the
aggregation transition is first-order like.
Note that the discontinuity of the aggregation transition has been also
shown for the simple off-lattice AB model without the directional hydrogen
bonds \cite{Janke06}.

\begin{figure}
\center
\includegraphics[width=3.4in]{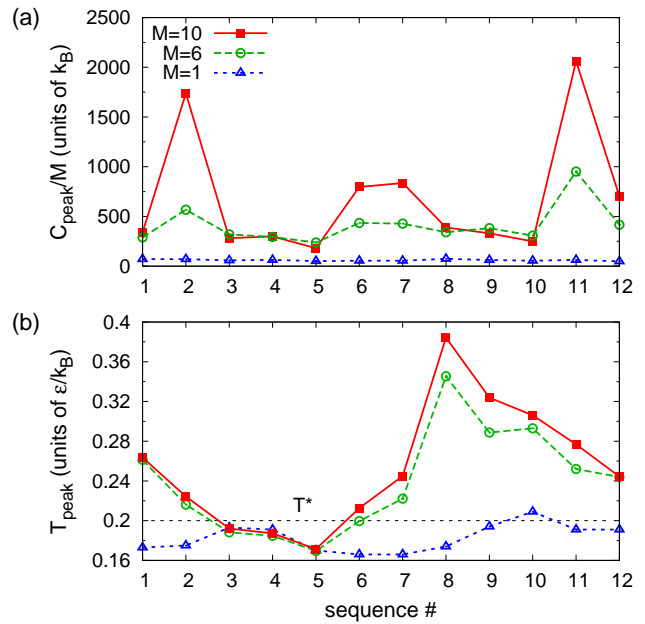}
\caption{\label{fig:cmax6}
Dependence of the maximum of the specific heat $C_\mathrm{peak}$ per molecule
(a) and its temperature $T_\mathrm{peak}$ (b) on the sequence for systems of
$M=10$ (solid), $M=6$ (dashed) and $M=1$ (dotted) peptides at 1 mM
concentration. The sequence number is given in accordance to Table I.  The
horizontal line in (b) indicates a putative physiological temperature $T^*$.
}
\end{figure}

We find that the specific heat strongly depends on both the sequence and the
system size. Fig. \ref{fig:s2} and Fig. \ref{fig:s4} show the temperature
dependence of the specific heat per molecule for various system sizes for
sequences S2 and S4, respectively. For sequence S2, the case in which
fibril-like aggregates form, it is shown that as $M$ increases the specific
heat's peak shifts toward higher temperature and its height increases
(Fig. \ref{fig:s2}). This result indicates that the
aggregate becomes increasingly stable and the transition becomes more
cooperative as the system size increases. The increasing cooperativeness
of the aggregation transition correlates with the increasing orderness in
the structure of the aggregate. For sequence S4, for which the aggregates are
helix bundles, the height of the main peak increases with $M$ but the position
of the peak varies non-monotonically (Fig. \ref{fig:s4}). Note that
the aggregation transition for sequences S4 is always found at a slightly lower
temperature than the folding transition of individual chain. This is in
contrast with sequence S2, whose aggregation transition temperature
is always higher than the folding temperature of a single chain.

In Fig. \ref{fig:cmax6}, the results of the maximum specific heat
per molecule, $C_\mathrm{peak}/M$, and the temperature of the peak,
$T_\mathrm{peak}$, are combined for all sequences considered and for several
values of $M$. It is shown that the variation of both $C_\mathrm{peak}/M$
and $T_\mathrm{peak}$ increases with $M$. Note that for $M=10$, the highest
specific heat maxima correspond to sequences S2 and S11 whose aggregates
are fibril-like (see Fig. \ref{fig:galm10}). 
Apart from the absolute value of $C_\mathrm{peak}$, the increase of
$C_\mathrm{peak}/M$ with $M$ is also a signature of cooperativity.
For sequences S2 and S11, $C_\mathrm{peak}/M$ is not only the highest
among all sequences but also increases with $M$ much faster than other
sequences, suggesting that these sequences have the most cooperative
aggregation transitions. Our results indicates that the propensity of forming
fibril-like aggregates is associated with the cooperativity of the aggregation
transition.

The wide variation in the transition temperatures $T_\mathrm{peak}$ among
sequences, as shown in Fig. \ref{fig:cmax6}b, suggests another interesting
aspect of aggregation. Suppose that we consider
the systems at the physiological temperature, $T^*$. In our model, a rough
estimate of $T^*$ could be 0.2 $\epsilon/k_B$, which corresponds to a local
hydrogen bond energy of 5 $k_BT^*$. For $M=1$, one finds that all sequences
but S10 has $T_\mathrm{peak} < T^*$ suggesting that the peptides are
substantially unstructured at $T^*$ as a single chain. For $M=6$ and $M=10$,
only three sequences, S3, S4 and S5, have $T_\mathrm{peak} < T^*$, while the
other have $T_\mathrm{peak} > T^*$. Thus, sequences S3, S4 and S5 do not
aggregate at $T^*$ while other sequences do. This result indicates that the
variation of aggregation transition temperatures among sequences is also a
reason why protein sequences behave differently towards aggregation at the
physiological temperature. Some sequences do not aggregate because aggregation
is thermodynamically unfavorable at this temperature.

Note that the ability of forming fibril-like aggregates is not necessarily
associated with a high aggregation transition temperature. In fact, Fig.
\ref{fig:cmax6}b shows that sequences S2 and S11 have only a medium value of
$T_\mathrm{peak}$ among all sequences, for both $M=6$ and $M=10$. Some
sequences with a higher $T_\mathrm{peak}$, such as S8, S9 and S10, form
disordered aggregates.

\begin{figure}
\center
\includegraphics[width=3.4in]{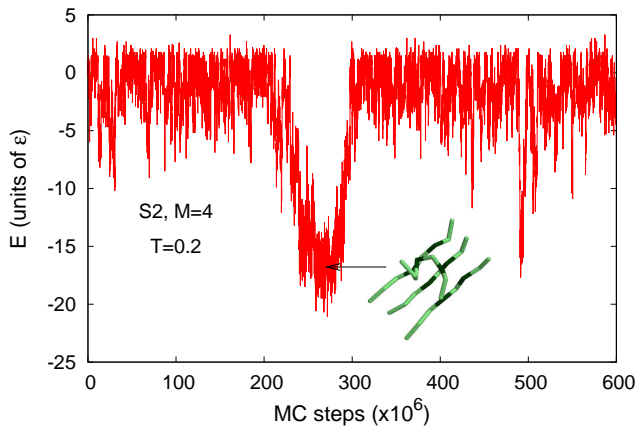}
\caption{\label{fig:s2m4}
Energy as function of Monte Carlo steps in a trajectory at $T=0.2$ for the
sequence S2 system with $M=4$ at 1 mM concentration. The conformation
shown is a metastable state with a 3-peptide $\beta$-sheet in contact with a
disordered helix formed by the 4th peptide.
}
\end{figure}

The dependence of specific heat on the system size also reveals a condition
for aggregation. Fig. \ref{fig:s2} shows that for sequence S2, systems of $M
\leq 4$ have the specific heat peaked at a lower temperature than
$T^*=0.2\epsilon/k_B$, which means that these systems do not aggregate at
$T^*$. Only for $M > 4$, the specific heat peak temperature is higher than
$T^*$ indicating that the fibril-like aggregates formed by this sequence are
stable at $T^*$. Thus, a sufficient number of peptides is needed for the
aggregation to happen at a given temperature. 
We also find that the lower peak in the specific heat of the system of $M=4$
(Fig. \ref{fig:s2}) corresponds to a transition from metastable aggregates at
intermediate temperature to the ground state at low temperature.
Fig. \ref{fig:s2m4} shows the trajectory of an equilibrium simulation at
$T=0.2\epsilon/k_B$ for sequences S2 with $M=4$. 
The time dependence of the system's energy in this trajectory indicates that the
peptides do not aggregate most of the time, so that the energy is relatively
high, but for some short periods they can spontaneously form a 
metastable aggregate of a much lower energy. This metastable aggregate has
a three-stranded $\beta$-sheet (Fig. \ref{fig:s2m4}, inset) and
could act as a template for fibril growth in systems of more peptides.

\subsection{Kinetics of fibril formation}

It is well-established that amyloid fibril formation follows the
nucleation-growth mechanism, familiar to that found in studies of crystallization
and polymer growth \cite{Oosawa1962}. The time dependence of fibril mass is
characterized by an initial lag phase, during which the growth rate is small,
before a period of rapid growth, resulting in sigmoidal kinetics
\cite{Radford2008,Hellstrand2009,Knowles2009}. Nucleation gives rise to the lag
phase and is a rate-limiting step.  
A primary nucleation event corresponds to the initial formation
of an amyloid-like aggregate from soluble species, which is followed by an
elongation of the fibrils through the templated addition of species.
Analyses of experimental kinetic data using master equation
indicate that amyloid fibril growth can be dominated by secondary nucleation
events such as fragmentation \cite{Knowles2009} and surface-catalyzed
nucleation \cite{Ruschak2007}. 
The nucleated and templated polymerization properties of fibril formation
have been shown in coarse-grained
\cite{Shea07,Nguyen2004,Nguyen2005,Auer2007,AuerPRL08} and all-atom
\cite{Hills2007} simulations of short peptides.  Studies of crystal-based
lattice models by using classical nucleation theory
\cite{Kashchiev2010,Auer2014} and simulations
\cite{Muthukumar2009,Irback2013} provide characterizations of the
nucleation barriers in terms of $\beta$-sheet growth within a layer and
intersheet couplings, together with extensive temperature and concentration
dependence.

In the following, we will investigate the behavior of fibril growth within our
tube model for sequence S2. Since the ground state for this sequence
is a two-layered $\beta$-sheet structure, we do not expect it to display very
rich behavior, such as the increase of fibril thickness by multi-step
$\beta$-sheet layer addition. Nevertheless, the system may be useful
for understanding the formation of a single protofilament.

\begin{figure}
\center
\includegraphics[width=3.4in]{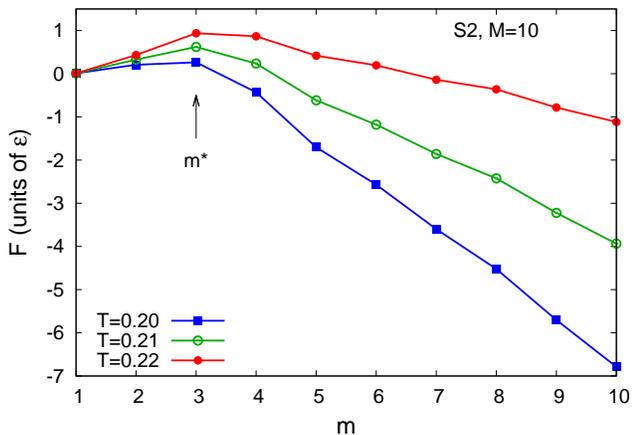}
\caption{\label{fig:fnmax}
Dependence of total free energy, $F$, on the size of the largest
aggregate, $m$, for the sequence S2 system of $M=10$ peptides 
at 1 mM concentration and at three
different temperatures, $T=0.2$, 0.21 and 0.22 $\epsilon/k_B$, as indicated.
The free energy of non-aggregated state, of $m=1$, is used as reference.
A barrier with the maximum located at $m=3$ is indicated.
}
\end{figure}

First, we consider a system of $M=10$ peptides with concentration $c=1$ mM
under equilibrium condition. Fig. \ref{fig:fnmax} shows the dependence of the
total free energy of the system on the size of the largest aggregate, $m$,
formed at three temperatures slightly below $T_\mathrm{peak}$ including
$T=T^*=0.2\,\epsilon/k_B$. 
This free energy is defined as $F(m,T) = - k_B T \log P(m,T)$, where
$P(m,T)$ is the probability of observing a conformation with the largest
aggregate size equal to $m$ at temperature $T$.
$P(m,T)$ was determined from parallel tempering simulations with the
weighted histogram method \cite{Ferrenberg1989}.
It is shown that for all these temperatures the free
energy has a maximum at $m=3$, suggesting that $m=3$ could be the size of the
critical nucleus for fibril formation. Interestingly, $M=3$ is also the system
size at which the ground state changes from a helix bundle to a $\beta$-sheet
on increasing $M$, and this $\beta$-sheet is unstable at temperatures larger or
equal $T^*$ (see Fig.  \ref{fig:s2}). Thus, there is a consistency between the
equilibrium data obtained with a small and a larger $M$ in terms of aggregation
properties.  The free energy barrier for aggregation in Fig. \ref{fig:fnmax} is
found to increase with $T$ and is about of $1\,k_B T$ to $4\,k_B T$.  This barrier is
not large and is consistent with the fact that the sequence considered is
highly aggregation-prone.  For $m > 3$, Fig. \ref{fig:fnmax} shows that the
free energy decreases almost linearly with $n$, which is consistent with the
fact that the growth of the aggregate in size is essentially one-dimensional.
After a certain size, new peptides join an existing aggregate from either of
its two ends and establish the elongation of the $\beta$-sheets.

\begin{figure*}
\center
\includegraphics[width=\textwidth]{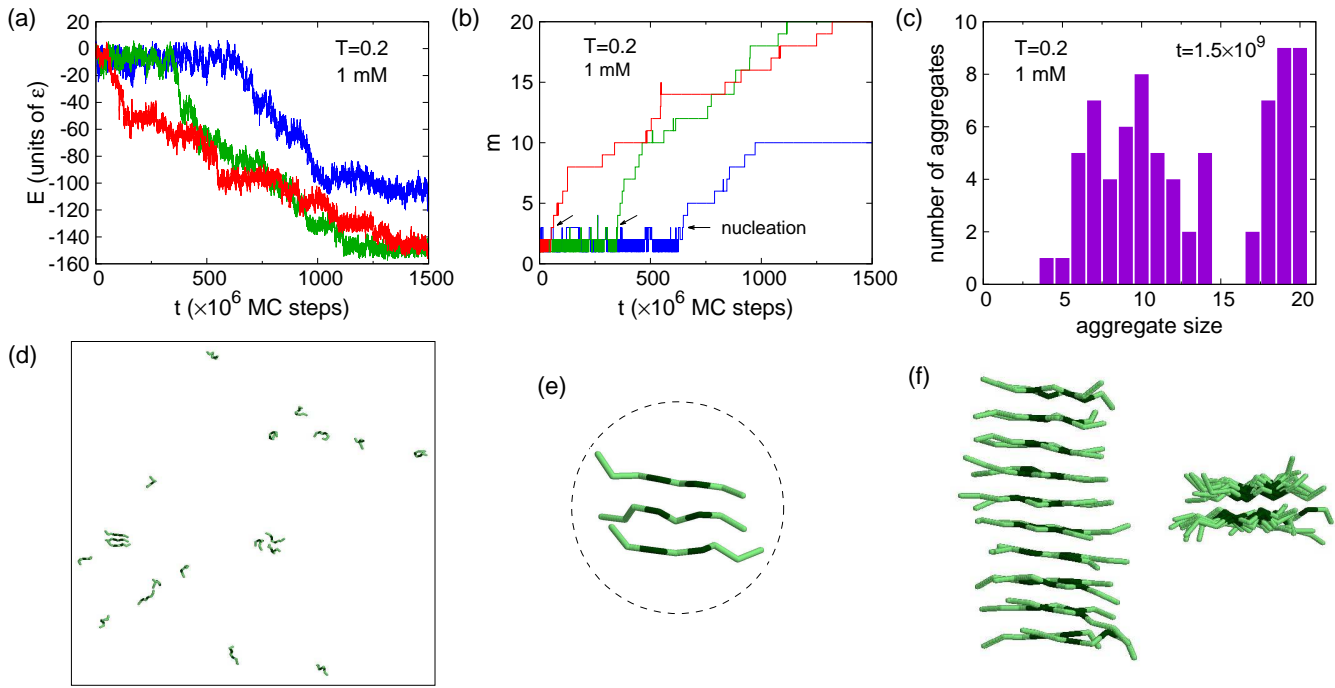}
\caption{\label{fig:s2m20}
Kinetics of fibril formation for sequence S2 with $M=20$ peptides at
concentration 1 mM and temperature $T=0.2 \epsilon/k_B$. (a) Dependence of the
energy, $E$, on time, $t$, measured in MC steps for three different
trajectories.  (b) Time dependence of the maximum aggregate size $m$ for the
same three trajectories as shown in (a). Arrows indicate nucleation event for
each trajectory.  (c) Histogram of the aggregate size given by the number of
peptides obtained at a large time of $t=1.5\times10^9$ MC steps.  (d) Snapshot
of peptide configuration at a nucleation event.  (e) Conformation of the
nucleated cluster formed by three peptides taken from the configuration shown
in (d). (f) Conformation of an elongated fibril-like structure formed by 20
peptides.
}
\end{figure*}

We then considered a larger system of $M=20$ peptides and studied the time
evolutions from random configurations of dispersed monomers. Up to 100
independent trajectories are carried out to determine the statistics.  We first
consider the system at concentration $c=1$ mM and $T=0.2\,\epsilon/k_B$.  Fig.
\ref{fig:s2m20} (a and b) shows three typical trajectories with the total
energy $E$ and the size of the largest aggregate $m$ as functions of time.
Interestingly, these trajectories show clear evidence of an initial lag time,
during which $m$ fluctuates but remains small ($m \leq 3$) before a rapid and
almost monotonic growth (Fig. \ref{fig:s2m20} b). They also shows that
nucleation is complete for $m=3$, in consistency with the equilibrium analysis
obtained before for $M=10$.  A peptide configuration at a nucleation event is
shown on Fig. \ref{fig:s2m20}d indicating that a possible nucleus is a
three-stranded $\beta$-sheet formed by three peptides (Fig. \ref{fig:s2m20}e).
Fig. \ref{fig:s2m20}c shows that the system can form multiple aggregates of
various sizes. The distribution of the aggregate size obtained after a
sufficient long time is bimodal reflecting the fact that the system size is
finite and clusters of less than 4 peptides are unstable. Thus, one either
observes one large cluster with size close to the system size or several
smaller clusters. The largest aggregates of $m=20$ peptides have the form of an
elongated double $\beta$-sheet strongly resemble a cross-$\beta$-structure
(Fig. \ref{fig:s2m20}f).

\begin{figure}
\center
\includegraphics[width=3.4in]{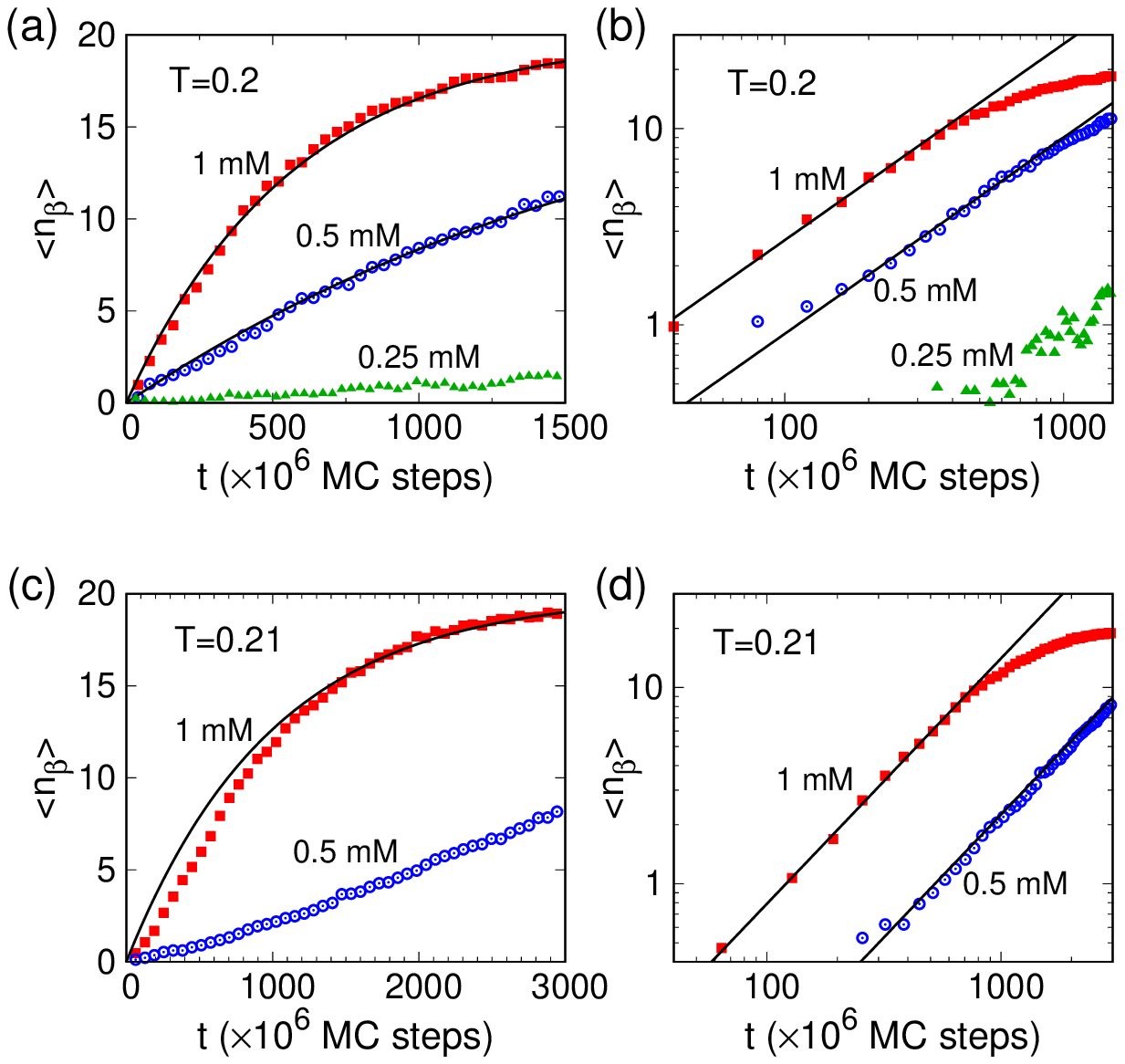}
\caption{\label{fig:nbeta}
Time dependence of the average number of peptides in $\beta$-sheet
conformation, $\langle n_\beta \rangle$, in the aggregation of sequence S2
with $M=20$. The system is considered at temperatures $T=0.2 \epsilon/k_B$
(a,b) and $0.21 \epsilon/k_B$ (c,d) and at several concentrations, $c = $ 1 mM
(squares), 0.5 mM (circles) and 0.25 mM (triangles), as indicated.  The average
of $n_\beta$ for each concentration is taken over 100 independent trajectories.
Right figures (b and d) plot the same data as in the left figures (a and c),
respectively, except that in log-log scale.  Data points are fitted to an
exponential relaxation function of $M(1 - e^{-t/t_0})$ for $c=1$ mM (solid) 
with $t_0=570\times 10^6$ for $c=1$ mM in (a) and $t_0=1850 \times 10^6$ for
$c=0.5$ mM in (a), and  $t_0= 10^9$ for $c=1$ mM in (c). The log-log
plots shows that the growth of $n_\beta$ at small times follows a power law,
$\langle n_\beta \rangle \propto t^\alpha$, with $\alpha=1$ in (b) and
$\alpha=1.25$ in (d) for both concentrations of 1 mM and 0.5 mM.
}
\end{figure}

Consider now the number of peptides in $\beta$-sheet conformation, $n_\beta$,
which counts all the peptides that have at least 4 consecutive hydrogen bonds
with another peptide. Fig. \ref{fig:nbeta} shows the dependence of $n_\beta$ on
time $t$, with $t$ measured in number of MC steps, averaged over the
trajectories, for two different temperatures and for various concentrations. It
is shown in Fig. \ref{fig:nbeta} (a and b) that for $T=0.2\,\epsilon/k_B$, the
time dependence of $\langle n_\beta \rangle$ can be fitted well to the
exponential relaxation function of $M(1-e^{-t/t_0})$, where $t_0$ is the
characteristic time of aggregation. This time dependence also depends strongly
on the concentration $c$ with $t_0$ increases more than 3 times by changing $c$
from 1 mM to 0.5 mM.  There seems to be no evidence of a lag phase at
$T=0.2\,\epsilon/k_B$ as $\langle n_\beta \rangle$ increases linearly with $t$
for small $t$ (Fig.  \ref{fig:nbeta}b). This lack of  evidence, however,
may be due to the fact that the deviation from the exponential growth is
too small to be observed. Indeed, we find that if the temperature is increased
a little to $T=0.21\,\epsilon/k_B$, the lag phase can be observed.  Fig.
\ref{fig:nbeta}c shows that the growth of $\langle n_\beta \rangle$
in time is significantly deviated from the exponential relaxation function at
small time. This growth when plotted in a log-log scale (Fig. \ref{fig:nbeta}c)
shows that at small time $\langle n_\beta \rangle \propto t^{\alpha}$ with
$\alpha \approx 1.25$. The exponent $\alpha > 1$ indicates that the time
dependence of $\langle n_\beta \rangle$ behaves like a convex function, which
proves the existence of the lag phase at small time. The stronger evidence of
the lag phase at $T=0.21\,\epsilon/k_B$ compared to that at
$T=0.2\,\epsilon/k_B$ is consistent with the higher free energy barrier for
nucleation at the former temperature previously shown in Fig. \ref{fig:fnmax}.
Note that the lag phase has been also observed in the aggregation of
homopolymers with a similar model but for a larger system \cite{Auer2007}.

With the limited system size and time scale considered, we have not observed
fragmentation of the fibril-like aggregates. On the other hand, the
surface-catalyzed nucleation may exist from perspective of
a two-layer $\beta$-sheet structure. The exposed hydrophobic side chains
of the nucleated three-stranded $\beta$-sheet promotes association of other
peptides by hydrophobic attraction. We find that clusters of 4 to 6
peptides often transform into a double $\beta$-sheet structure before
continuing to grow. Thus, this secondary nucleation is surface-catalyzed
and follows immediately after the primary nucleation event.
The secondary nucleation also helps to stabilize the primary nucleus.

\begin{figure}
\center
\includegraphics[width=3.4in]{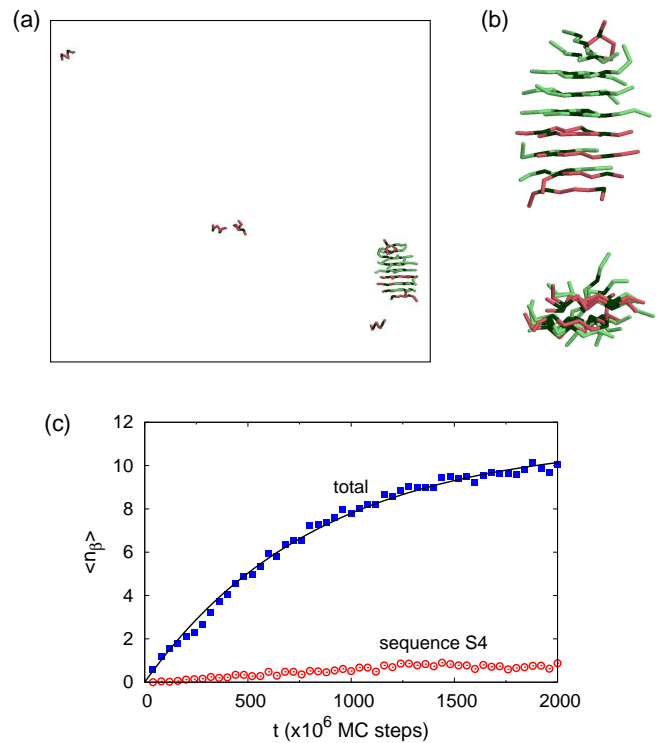}
\caption{\label{fig:mix}
(a) Snapshot of a conformation obtained in a simulation of the binary mixture of
10 chains of sequence S2 and 10 chains of sequence S4 at concentration $c=1$ mM
and temperature $T=0.2\epsilon/k_B$. H residues are shown in dark green. P
residues are in light green and pink colors for the S2 and S4 chains,
respectively. (b) Zoom-in side and top views of the aggregate shown in a. Note
that six S4 chains are present in the aggregate, and five of them are in the
$\beta$-sheet configuration.
(c) Time dependence of the average number of peptides in $\beta$-sheet conformation,
$\langle n_\beta \rangle$, obtained from 100 independent simulations, for both
sequences together (squares) and for sequences S4 only (circles). A fit to the
exponential relaxation function as given in the caption of Fig. \ref{fig:nbeta}
with $t_0=832\times 10^6$ (solid line) is shown for the case of both sequences.
}
\end{figure}

\subsection{Aggregation of mixed sequences}

Finally, we study the aggregation for a binary mixture of two sequences, S2 and
S4. It was shown that in homogeneous systems, the first sequence is strongly
fibril-prone, whereas the second one forms only $\alpha$-helices. Furthermore,
the sequence S4 has the aggregation transition temperature lower than $T^*$, so
the its aggregate is not stable at $T^*$. Strikingly, our simulations at $T^*$
show that in a binary system of equally 10 chains of each sequence, after a
sufficiently long time, a fraction of the S4 chains aggregate and convert into
$\beta$-sheet conformation on an existing aggregate formed by the S2 chains
(see Fig. \ref{fig:mix}). Though this fraction is only about 10\% on average,
this observation shows that the template-based mechanism for fibril formation
can be effective for polypeptides of very different natures. Here, the
fibril-like aggregate formed by the aggregation-prone peptides acts as the
template for the aggregation of non-aggregation-prone peptides.
Note that due to the mismatch of different hydrophobic patterns in the two
sequences, the aggregates formed by the two sequences are more disordered than
the homogeneous ones (Fig. \ref{fig:mix}b). It is also shown in Fig.
\ref{fig:mix}c that the growth of this mixed aggregate at the given temperature
remains exponential but the characteristic time for aggregation is
larger than in corresponding homogeneous system of sequence S2.

\section{Discussion}

Previous study of the tube model \cite{HoangPNAS06} has shown that
hydrophobic-polar sequence can select protein's secondary and
tertiary structures. In particular, the HPPH and HPPPH patterns have been
identified as strong $\alpha$-formers, whereas the HPH pattern is a
$\beta$-former. Strikingly, exactly the same binary patterns have been used in
experiments that allow the successful design of de novo proteins
\cite{Hecht1993,Hecht2003}.  In the present study, we find that these simple
selection rules still hold for the peptides in aggregates, even though the
model has been changed by considering the orientations of side chains. The
present study shows that the binary pattern also determines the orderness of
the aggregate. In particular, there should be some compatibility between
the alignment of hydrophobic side chains and the overall symmetry of
the aggregate. Interestingly, the HPH pattern appears to be both a strong
$\beta$-former and a highly aggregation-prone sequence. Our finding is in a
full agreement with experimental design of amyloids \cite{Hecht1999}, which
shows that segments of alternating hydrophobic and polar pattern (such as
PHPHPHP) can direct protein sequences to form amyloid-like fibrils. The effect
of this pattern has been also reported in simulations of an
off-lattice model \cite{Shea07} and also in a recent study of a lattice
model \cite{Abeln2014}. Interesting, it has been found that
Nature disfavors this pattern in natural proteins \cite{Hecht1999}.

The role of side-chains in amyloid fibril formation has been stressed in
early all-atom simulations of short peptides. The study by Gsponer {\it et al.}
\cite{Gsponer2003} showed that backbone hydrogen bonds favor the antiparallel
$\beta$-sheet packing but side-chain interactions stabilize the in-register
parallel $\beta$-sheet aggregate. The simulations performed by de la Paz {\it
et al.} \cite{Colombo2005} indicated the importance of specific contacts among
side-chains at specific sequence position for the formation and stabilization
of $\beta$-sheet oligomers and ordered fibrils. The exclude volume of
side-chains alone has been show to enhance the formation of helices
\cite{BanavarPNAS09} and planar sheets \cite{Skrbic2016}. 
A recent lattice model showing the formation of ordered fibrils includes the
side chain directionality \cite{Abeln2014}.
Here, we show that the correlated orientations of hydrophobic side-chains are
important for both the ordered packing of $\beta$-strands within a
$\beta$-sheet and the stacking of $\beta$-sheets in the fibril. In particular,
the alternating hydrophobic polar pattern leads to $\beta$-sheets of
hydrophobic side chains oriented on one side of the $\beta$-sheet. 
This one-sided orientation stabilizes the two-layered $\beta$-sheet aggregate,
which is the system's ground state and can grow into a long fibril, as
shown for the case of sequence S2. Note that the asymmetry of hydrophobic
$\beta$-sheet surfaces has been considered in a lattice model \cite{Auer2014},
showing increased stabilities of multi-layered $\beta$-sheets that have weak
hydrophobic surfaces exposed. Our study shows how this asymmetry is induced by
the sequence at molecular level. 

Previous studies \cite{Auer2010,Auer2011} have indicated that few-layered
$\beta$-sheet aggregates can be stable with respect to the peptide solution and
to liquid-like oligomers in certain ranges of temperature and concentration,
but are metastable with respect to the aggregates of large and infinite number
of $\beta$-sheet layers. The example given by our sequence S2 shows that it is
possible to design a thermodynamically stable fibril of a fixed small number of
$\beta$-sheet layers by using appropriate amino acid sequences. This result is
supported by the common observation of the finite and rather uniform thickness
of amyloid fibrils, even though some short peptides are reported to form
nanocrystals \cite{Griffin2007} at low peptide concentrations.

Our thermodynamics calculations show that the formation of fibril-like
aggregates is much more cooperative than that of non-fibril-like aggregates.
This cooperativity was indicated by both the height of the specific heat peak
and the increase of the maximum specific heat per molecule with the system
size. The high cooperativity of fibril formation can be understood as due to
the highly ordered nature of fibril structures and the dominating contribution
of intermolecular interactions in these structures.  We also find that
thermodynamic stability is not a distinguished feature of fibril-like
aggregates. In particular, sequences associated with very high aggregation
transition temperature do not necessarily form fibril-like aggregates.
The increased overall hydrophobicity of the sequence is shown to enhance
the stability of the aggregates without impact on their
fibril characteristics.

It has been suggested \cite{Auer2007} that on increasing peptide concentration or
peptide hydrophobicity, amyloid fibril nucleation changes from one-step, i.e. the
ordered nucleus is formed directly by monomeric peptides from the solution, to
the two-step condensation-ordering mechanism, in which nucleation is
preceded by the formation of large disordered oligomers. It has been also
shown that the nucleation pathway depends on the sequence and its
hydrophobicity \cite{Luiken2015}.
The sequence S2 in our study shows the one-step nucleation, consistent with the
scenario suggested in Ref. \cite{Auer2007}, given that this sequence has a
relatively low hydrophobicity and the 1 mM concentration considered in the
simulations is not high compared to those considered in Ref. \cite{Auer2007}.
The impact of the HP sequence on nucleation is also associated with a small
nucleation barrier and the rapid nucleation with almost invisible lag phase
observed for this sequence. 
For this fibril-prone sequence, it is found that the non-equilibrium behavior of a
larger system is consistent with equilibrium properties of smaller systems at
the same peptide concentration. In particular, the frequent formation and
dissolving of the aggregates before nucleation and the growth of the
aggregates after nucleation are in accord with their thermodynamic
stabilities as isolated systems. Note that in general, fibril formation can be
kinetics dependent \cite{Auer2012B} rather than thermodynamics, especially at
very low or very high concentrations. 
Interestingly, the small size of the critical nucleus found in our study agrees
with those obtained in homopolymer studies \cite{Auer2007,AuerPRL08} as well as
in lattice heteropolymer \cite{Co2012} and all-atom \cite{Hills2007,Nguyen2007}
simulations of short peptides.

In a recent experiment, Ridgley {\it et al.} \cite{Ridgley2011} show that
mixtures of aggregation-prone peptides and proteins, including the rich in
$\alpha$-helices myoglobin, self-assemble into amyloid fibers with increased
amounts of cross-$\beta$ content. It was suggested that the $\beta$-sheet
template formed by the peptides promotes the $\alpha$ to $\beta$ conversion in
the proteins and their involvement in the cross-$\beta$ structure. Our
simulation result on the peptide binary mixture is fully consistent with
this experiment and shows that a cross-$\beta$-sheet can be heterogeneous in
its peptide composition. It is possible that naturally occurring amyloid
fibrils can possess this heterogeneity due to the templated self-assembly
process. A certain degree of heterogeneity can be seen in the fibril
structure of HET-s prion protein \cite{Meier2008}, which shows that
the cross-$\beta$-sheets are formed by repeating `in-register' protein segments
but neighboring $\beta$-strands do not have the same amino acid sequence.

\section{Conclusion}

The present study has highlighted several aspects of amyloid fibril formation
that include the sequence determination of fibrillar structures, the role of
side chain directionality, the thermodynamics of aggregation, and the
nucleation and template-based growth mechanism. In agreement with various
experimental findings, our results indicate that fibril-like aggregates form
very much under the same principles as in protein folding, such as the
alignment of hydrophobic residues in a $\beta$-sheet, the packing of
hydrophobic side chains, and the cooperativity of the aggregation transition.
These principles are mainly associated to the specificity of a sequence. Our
simulations also show another feature of amyloid formation, that is
considerably non-specific to a sequence, namely the fibril-induced aggregation
of a non-aggregation-prone sequence. This templating property certainly
complicates the problem of amyloid formation as it suggests that the
cross-$\beta$-structure can be heterogeneous in their sequence or peptide
composition. Our study provides a basis for finding the routes to deal with the
problem.

This research is funded by Vietnam National Foundation for Science and
Technology Development (NAFOSTED) under Grant No. 103.01-2016.61. 
The use of computer cluster at CIC-VAST is gratefully acknowledged.

\bibliography{refs_amyloid}

\end{document}